\documentstyle[11pt,newpasp,twoside,epsf]{article}
\def\edcomment#1{\iffalse\marginpar{\raggedright\sl#1\/}\else\relax\fi}
\reversemarginpar

\begin{document}
\title{Theory of Microlensing}
 \author{Andrew Gould}
\affil{Dept.\ of Astronomy, Ohio State University, 140 W.\ 18th Ave.,
Columbus, OH, 43210-1173}

\begin{abstract}

	I present a somewhat selective review of microlensing theory, 
covering five major areas: 1) the derivation of the basic formulae,
2) the relation between the observables and the fundamental physical
parameters, 3) binaries, 4) astrometric microlensing, and 5) femtolensing.

\end{abstract}

\section{Introduction}

	All of gravitational microlensing is reducible to a single
equation, the Einstein formula for $\alpha$, the deflection
of light from a distant source passing 
by a lens of mass $M$ at an impact parameter $b$, 
\begin{equation}
\alpha = {4 G M\over b c^2}.
\end{equation}
This equation has been verified experimentally by Hipparcos to an accuracy
of 0.3\% (Froeschle, Mignard, \& Arenou 1997).
Despite its apparent simplicity, equation (1) generates an incredibly
rich phenomenology.  The aim of this review is to present the reader
with a concise introduction to microlensing phenomena from the standpoint
of theory.  It is impossible to cover all aspects of microlensing in the
space permitted.  Several good reviews of microlensing can be consulted
to obtain a deeper appreciation for various aspects of the subject
(Paczy\'nski 1996; Gould 1996; Roulet \& Mollerach 1997; Mao 1999b).

	By definition, microlensing is gravitational lensing where the
images are too close to be separately resolved.  The main microlensing
effect that has been discussed in the literature (and the only one
that has actually been observed) is {\it photometric microlensing}: 
the magnification of the source due to the convex nature of the lens
(Einstein 1936; Refsdal 1964; Paczy\'nski 1986; Griest 1991; Nemiroff 1991).  
However, there are two other effects that deserve attention from the 
standpoint of theory: {\it astrometric microlensing}, the motion of the
centroid of the images relative to the source, and {\it femtolensing},
which refers to interference effects in microlensing.

\section{Observables and Physical Parameters}

\subsection{The Lens Diagram}

\begin{figure}[h]
\epsfxsize \hsize
\epsfbox{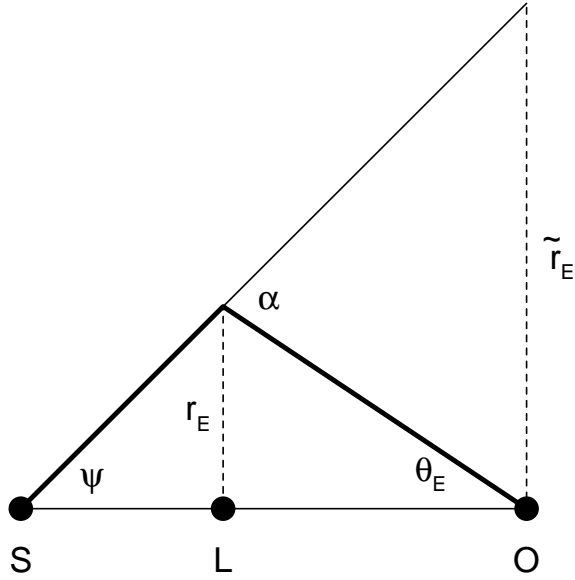}
\caption{Basic geometry of lensing for the case when the source (S), lens (L),
and observer (O), are aligned.  The light is deflected by an angle $\alpha$
into a ring of radius $\theta_{\rm E}$.  The resulting Einstein ring, 
$r_{\rm E}$ projected onto the observer plane is $\tilde r_{\rm E}$.
Simple geometry relates the observables,  $\theta_{\rm E}$ and
$\tilde r_{\rm E}$ to the lens mass $M$ and lens-source relative paralax
$\pi_{\rm rel}$.  See eqs.\ (2) and (3).}
%\caption{la la}
\end{figure}

	Consider a lens at $d_l$ and a more distant source at $d_s$
that are perfectly coaligned with the observer (Fig.\ 1).  By axial symmetry,
the source is imaged into a ring (the ``Einstein ring'') whose angular
radius is denoted $\theta_{\rm E}$.  The impact parameter, $r_{\rm E}$, is
called the ``physical Einstein ring'', and its projection back onto the plane
of the observer is called the ``projected Einstein ring'', $\tilde r_{\rm E}$.
As I will show below, $\theta_{\rm E}$ and  $\tilde r_{\rm E}$, are two
of the seven observables of the system.  Using the small-angle approximation,
they can be related to the physical parameters, $M$ and the 
source-lens relative parallax, $\pi_{\rm rel}$ (Gould 2000b).  First, 
$\alpha/\tilde r_{\rm E} = \theta_{\rm E}/r_{\rm E}$, so using equation
(1), one finds
\begin{equation}
\tilde r_{\rm E}\theta_{\rm E} = \alpha r_{\rm E} = {4 G M\over c^2}.
\end{equation}
Second, using the exterior angle theorem, $\theta_{\rm E} = \alpha-\psi
= \tilde r_{\rm E}/d_l - \tilde r_{\rm E}/d_s$, so
\begin{equation}
{\theta_{\rm E}\over \tilde r_{\rm E}} = {\pi_{\rm rel}\over \rm AU}.
\end{equation}
These equations can easily be combined to yield,
\begin{equation}
\theta_{\rm E} = \sqrt{{4 G M\over c^2}\,{\pi_{\rm rel}\over \rm AU}}, \qquad
\tilde r_{\rm E} = \sqrt{{4 G M\over c^2}\,{\rm AU \over \pi_{\rm rel}}}.
\end{equation}
Note that if $\theta_{\rm E}$ and $\pi_{\rm rel}$ are measured in mas,
$\tilde r_{\rm E}$ is measured in AU, and $M$ is measured in $M_\odot/8$,
then all numerical factors and physical constants in these last three
equations can be ignored.

\subsection{The Lensing Event}

	If the lens is not perfectly aligned with the source, then the
axial symmetry is broken and there are only two images.  A similar use
of the exterior-angle theorem then yields 
$\theta_I^2 - \theta_I\theta_s = \theta_{\rm E}^2$, for the relation between
the image positions, $\theta_{I,\pm}$ and the source position $\theta_s$,
relative to the lens.
It is conventional to normalize the (vector) source position to 
$\theta_{\rm E}$, i.e., ${\bf u}\equiv \vec\theta_s/\theta_{\rm E}$, so that
the image positions are at
\begin{equation}
\vec\theta_{I,\pm} = \pm u_\pm \hat{\bf u},\qquad u_\pm \equiv 
{\sqrt{u^2+4} \pm u\over 2}.
\end{equation}
	
\begin{figure}[h]
\epsfxsize \hsize
\epsfbox{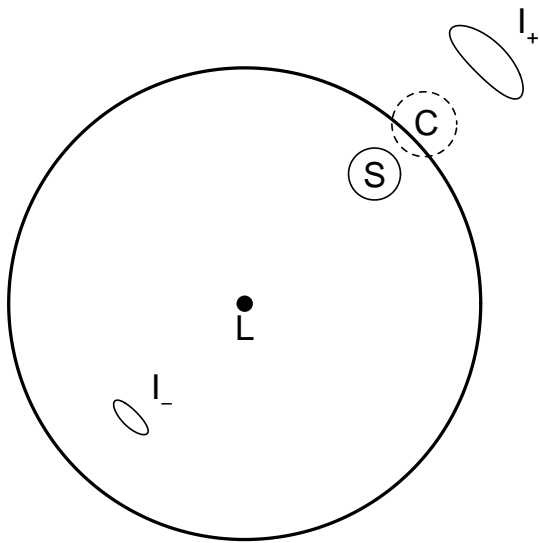}
\caption{Source and images for a point lens.  The bold line shows the Einstein
ring centered on the lens (L).  The two images (I$_+$ and I$_-$)
of the source (S) are shown with their correct relative size and shape.
The centroid of light (C) is shown at its correct position and with a size
proportional to the magnification but, since the centroid is by definition
unresolved, its shape is displayed arbitarily as a circle.}

\end{figure}

	By Liouville's theorem, surface brightness is conserved, so 
for a uniformly bright source, 
the magnification of each of the two images $A_\pm$ is given by the ratio
of the area of the image to the area of the source.  See Figure 2.
In the limit of a point
source, this ratio reduces to the Jacobian of the transformation,
\begin{equation}
A_\pm = \bigg|{\partial\vec\theta_{I,\pm}\over \partial\vec\theta_s}\bigg|
= {u_\pm^2\over u_+^2-u_-^2}% = {A\pm 1\over 2}
,\qquad A = A_+ + A_- =
{u^2+2\over u\sqrt{u^2+4}}
\end{equation}
Since $A$ is a monotonic function of $u$, the signature
of a {\it microlensing event} is that the source becomes brighter and then
fainter as the line of sight to the source gets closer to and farther from
the lens.  If the source, lens, and observer are all in rectilinear motion,
then by the Pythagorian theorem, $u(t)=[u_0^2 + (t-t_0)^2/t_{\rm E}^2]^{1/2}$,
where $t_0$ is the time of closest approach, $u_0=u(t_0)$, 
\begin{equation}
t_{\rm E}\equiv{\theta_{\rm E}\over\mu_{\rm rel}},\qquad
\vec \mu_{\rm rel} \equiv \vec\mu_l - \vec\mu_s,
\end{equation}
and $\vec\mu_{\rm rel}$ is the lens-source relative proper motion.

\subsection{Observables}

	From a photometric microlensing event, one can then usually 
measure three parameters, $t_0$, $u_0$, and $t_{\rm E}$.  Of these,
the first two tell us nothing about the lens, and the third is related
to $M$, $\pi_{\rm rel}$, and $\mu_{\rm rel}$ in
a complicated way through equations (4) and
(7).  However, as mentioned above, there are actually seven
(scalar) quantities that can in principle be observed in a microlensing
event.  These are: $t_{\rm E}$, $\theta_{\rm E}$, $\tilde r_{\rm E}$, 
$\phi$ (the angle of lens-source relative proper motion), and the source
parallax and proper motion,
$\pi_s$, and $\vec\mu_s$, which can be measured 
astrometrically after the event.

	To date, there have been only a handful of measurements of
$\theta_{\rm E}$, $\tilde r_{\rm E}$, and $\phi$, no measurements of
$\vec\mu_s$, and only estimates of $\pi_s$ (although these are probably very
good).  However,
all that could radically change in the next decade.  I first discuss
what it means that these quantities are ``observable'' and then briefly
outline future prospects.

	The angular Einstein ring $\theta_{\rm E}$ can be measured by
scaling the event against some known ``angular ruler'' on the sky.
The only such ``ruler'' to be used to date is the angular size of the source
(Gould 1994a; Nemiroff \& Wickramasinghe 1994; Witt \& Mao 1994),
which can be determined from the source flux and color, together with the 
empirical color/surface-brightness
relation (van Belle 1999; Albrow et al.\ 2000a).  So far, $\theta_{\rm E}$
has been measured for only 7 of the $\sim 500$ events detected to date 
(Alcock et al.\ 1997,2000; Albrow et al.\ 1999a,2000a; Afonso et al.\ 2000).
Many other methods have been proposed (see Gould 1996 for a review; 
Han \& Gould 1997), but none have been carried out.

	The projected Einstein ring $\tilde r_{\rm E}$ can be measured by
scaling the event against some known ``physical ruler'' in the plane of the
observer.  Three such ``rulers'' have been proposed, but the only one 
to be used to date is the Earth's orbit which induces a wobble
in the light curve (Gould 1992b).  Because this wobble depends on the 
Earth's motion being non-rectilinear, it is only significant if the event
lasts more than a radian (i.e., $t_{\rm E}\ga 58\,$days), and such events
are very rare.  To date, only about a half dozen events have measured 
$\tilde r_{\rm E}$
(Alcock et al.\ 1995; Bennett et al.\ 1997; Mao 1999a).  Another approach
is to observe the event simultaneously from two locations (``parallax'').
Since $\tilde r_{\rm E}\sim 5\,$AU, it would be best if the second location
were in solar orbit (Refsdal 1966; Gould 1994b).  

	The essential idea of a parallax satellite is illustrated in Figure 5 
of Gould (1996).  Basically, the Earth and the satellite each see a different
microlensing event characterized respectively by 
$(t_{0,\oplus},u_{0,\oplus},t_{\rm E,\oplus})$, and
$(t_{0,\rm sat},u_{0,\rm sat},t_{\rm E,\rm sat})$.  To
zeroth order $t_{\rm E,\rm sat}\simeq t_{\rm E,\oplus}$, but the 
differences in the other two components,
\begin{equation}
\Delta {\bf u} \equiv (\Delta t_0/t_{\rm E},\Delta u_0)
=(t_{0,\rm sat}/t_{\rm E},u_{0,\rm sat})-(t_{0,\oplus}/t_{\rm E},u_{0,\oplus}),
\end{equation}
give the displacement in the Einstein ring of the  satellite relative to
the Earth.  That is,
\begin{equation}
\tilde r_{\rm E} = {d_{\rm sat}\over \Delta u},
\end{equation}
where $d_{\rm sat}$ is the Earth-satellite separation (projected onto the
plane perpendicular to the line of sight).  There is actually a four-fold
ambiguity in $\Delta u_0 = \pm (u_{0,\rm sat}\pm u_{0,\oplus})$ but this
can be resolved, at least in principle, from the small difference in
Einstein timescales which constrains $\Delta u_0$ by (Gould 1995),
\begin{equation}
w_\perp \Delta u_0 + w_\parallel{\Delta t_{\rm E}\over t_{\rm E}}
+ d_{\rm sat}{\Delta t_{\rm E}\over t_{\rm E}^2}=0,
\end{equation}
where $w_\parallel$ and $w_\perp$ are the components of the Earth-satellite
relative velocity respectively parallel and perperdicular to the 
Earth-satellite separation vector.  Boutreux \& Gould (1996) and Gaudi \&
Gould (1997) showed that the degency could be broken with relatively modest
satellite parameters.  Unfortunately, no such satellite has been launched.  

	In a few specialized situations, it should
be possible to obtain parallaxes using Earth-sized baselines 
(Hardy \& Walker 1995; Holz \& Wald 1996; Gould 1997; Gould \& Andronov 1999;
Honma 1999), but to date no such measurements have been made.  All measurements
of $\tilde r_{\rm E}$ simultaneously measure $\phi$, and to date no other
measurements of $\phi$ have been made.

\subsection{Physical Parameters in Terms of Observables}

	It is convenient to replace $t_{\rm E}$ and $\phi$ together by
$\vec\mu_{\rm E}$ whose direction is $\phi$ and whose magnitude is
$\mu_{\rm E}\equiv t_{\rm E}^{-1}$, and to replace $\tilde r_{\rm E}$ by
$\pi_{\rm E}\equiv {\rm AU}/\tilde r_{\rm E}$.  Then equation (4) can be
rewritten as,
\begin{equation}
\theta_{\rm E} = \sqrt{\kappa M \pi_{\rm rel}},\qquad
\pi_{\rm E} = \sqrt{\pi_{\rm rel}\over \kappa M},\qquad 
\kappa\equiv {4 G \over c^2\, \rm AU}\simeq 8.144\,{{\rm mas}\over M_\odot},
\end{equation}
and the physical parameters
can be written in terms of the observables as,
\begin{equation}
M = %{c^2 {\rm AU}\over 4 G}\,
{\theta_{\rm E}\over \kappa\pi_{\rm E}},\qquad
\pi_l = \pi_{\rm E}\theta_{\rm E} + \pi_s,\qquad
\vec\mu_l = \vec\mu_{\rm E}\theta_{\rm E} + \vec\mu_s.
\end{equation}

\section{Astrometric Microlensing}

	While microlensing images cannot be resolved, the centroid of the
images deviates from the source position by
\begin{equation}
\delta\vec\theta = {A_+\vec\theta_{I,+} + A_-\vec\theta_{I,-}\over A_+ + A_-}
-\vec\theta_s = {{\bf u}\over u^2+2}\theta_{\rm E},
\end{equation}
which reaches a maximum of $\theta_{\rm E}/\sqrt{8}$ when $u=\sqrt{2}$.
See Figure 1.
Since typically $\theta_{\rm E}\sim 300\mu$as, $\delta\vec\theta$ is well
within the range of detection of the Space Interferometry Mission (SIM) and
perhaps ground-based interferometers as well.
While it is not obvious from equation (13), if ${\bf u}(t)$ is rectilinear,
then $\delta\vec\theta$ traces out an ellipse.  The size of the ellipse
gives $\theta_{\rm E}$ and its orientation gives $\phi$.  Hence, if 
microlensing were monitored astrometrically, it would be possible to
{\it routinely}  recover these two parameters 
(Boden, Shao, \& Van Buren 1998; Paczy\'nski 1998).  In fact, astrometric
microlensing has a host of other potential applications including
measurement of the lens brightness (Jeong, Han, \& Park 1999; 
Han \& Jeong 1999),
removal of degeneracies due to blended light from unmicrolensed sources
(Han \& Kim 1999), and
the detection and characterization of planetary
(Safizadeh, Dalal \& Griest 1999) and
binary microlenses
(Chang \& Han 1999; Han, Chun, \& Chang 1999).

	Because at late times the astrometric signature falls off as
$u^{-1}$ (eq.\ 13), compared to $u^{-4}$ for the photometric signature
(eq.\ 6), it could in principle be possible to measure $\tilde r_{\rm E}$
astrometrically from the Earth's orbital motion, even for events with
$t_{\rm E}\ll 58\,$days (Boden et al.\ 1998; Paczy\'nski 1998).  
If practical, this would mean that all the observables listed in \S\ 2.3
could be extracted from astrometric observations alone.  Unfortunately,
such measurements are not practical (Gould \& Salim 1999).

	Nevertheless, using SIM one can in fact extract all seven parameters,
and therefore can accurately determine both the masses and distances of 
the lenses.  SIM makes its astrometric measurements by centroiding the
fringe, i.e., by {\it counting photons} as a function of fringe position.
This means that SIM's {\it astrometric} measurements are simultaneously
{\it photometric} measurements.  Since SIM will be launched into solar
orbit, it therefore can effectively act as a parallax satellite (Gould \& 
Salim 1999).  Moreover, SIM can break the degeneracy in $\Delta u_0$ in two
ways: photometrically (according to eq.\ 10) and astrometrically
(by measuring the angle $\phi$ associated with the astrometric ellipse,
eq.\ 13).

	An alternate approach to measuring $\tilde r_{\rm E}$ would
be to compare SIM and ground-based {\it astrometry} rather than 
{\it photometry} (Han \& Kim 2000).

	Another important application of astrometric micolensing is
to measure the masses of nearby stars (Refsdal 1964; Paczy\'nski 1995, 1998;
Miralda-Escud\'e 1996).  Equation (13) still effectively describes the
astrometric deflection, but since typically $u\gg1$, this equation can
be more simply written as 
\begin{equation}
\Delta\theta = %{4 G \over c^2\rm AU}\,
{\kappa M\pi_{\rm rel}\over \theta_{\rm rel}},
\end{equation}
where $\theta_{\rm rel}\equiv |\vec\theta_l-\vec\theta_s|$.  In this form,
it is clear that the probability that a given lens will come close enough
to a background source to allow a mass measurement of fixed precision in
a fixed amount of time is
\begin{equation}
P \propto M\pi_l\mu_l N,
\end{equation}
where $N$ is the density of background sources (Gould 2000a).  Hence, the
best place to look for such candidates is a proper-motion selected catalog
near the Galactic plane.  In fact, the selection of such candidates is
a complex undertaking, but good progress is being made (Salim \& Gould 2000).

\section{Binary Lenses}

	Binary microlensing is one of the most active fields of theoretical
investigation in microlensing today.  In part this is due to the mathematical
complexity of the subject and in part to the demands that are being placed
on theory by new, very precise observations of binary events.  Schneider
\& Weiss (1986) made a careful early study of binary lenses despite the
fact that they never expected any to be detected (P.\ Schneider 1994,
private communication), in order to learn about caustics in quasar
macrolensing.  Indeed caustics are the main new features of binaries
relative to point lenses.  These are closed curves in the source plane
where a point source is infinitely magnified.  The curves are composed of
3 or more concave segments that meet at cusps.  Binary lenses can have
1, 2, or 3 closed caustic curves.  If the two masses are separated by 
approximately an Einstein radius, then there is a single 6-cusp caustic.
If they are separated by much more than an Einstein ring, then there are
two 4-cusp caustics, one associated with each member of the binary.  If the
masses are much closer than an Einstein ring, there is a central 4-cusp
caustic and two outlying 3-cusp caustics.  Figure 3 shows two cases of
the 6-cusp caustic, one close to breaking up into the two caustics 
characteristic of a wide binary and the other close breaking up into the three
caustics characteristic of a close binary.  Witt (1990) developed a simple
algorithm for finding these caustics.  Multiple-lens systems can have
even more complicated caustic structures (Rhie 1997; Gaudi, Naber, \& Sackett
1998).
\begin{figure}
\epsfxsize \hsize
\epsfbox{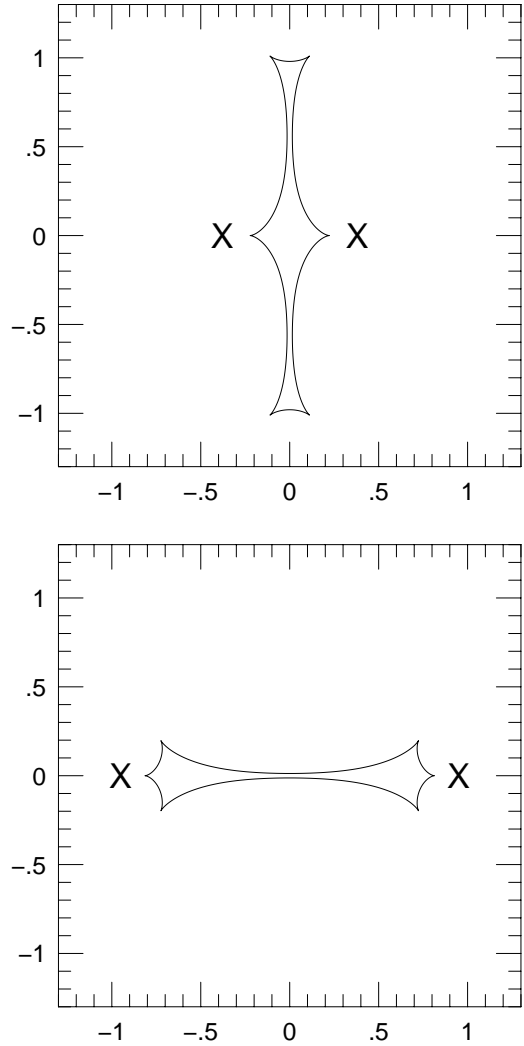}
\caption{Two extreme examples of 6-cusp caustics generated by
equal mass binaries.  The tick marks are in units of Einstein radii.
In each case, the crosses show the positions of the two components.
The upper panel shows a relatively close binary with the components
separated by $d=0.76$ Einstein radii.  For $d<2^{-1/2}$ the caustic would
break up into three caustics, a central 4-cusp caustic plus two outlying
3-cusp caustics.  The lower panel shows a relatively wide binary with $d=1.9$.
For $d>2$ the caustic would break up into two 4-cusp caustics.}
\end{figure}

\subsection{Binary Lens Parameters}

	Recall that a point-lens light curve is defined by just three
parameters, $t_0$, $u_0$, and $t_{\rm E}$.  These three generalize to
the case of binaries as follows: $u_0$ is now the smallest separation of
the source relative to the center of mass (alternatively geometric center)
of the binary, $t_0$ is the time when $u=u_0$, and $t_{\rm E}$ is the
timescale associated with the combined mass of the binary.  At least three
additional parameters are required to describe a binary lens: the angle
$\alpha$ at which the source crosses the binary axis, the binary mass ratio
$q$, and the projected separation of the binary in units of the Einstein ring.
Several additional parameters may be required in particular cases. If 
caustic crossings are observed, then the infinite magnification of the caustic
is smeared out by the finite size of the source, so one must specify
$\rho_*=\theta_*/\theta_{\rm E}$, where $\theta_*$ is the angular size of
the source.  If the observations of the crossing are sufficiently precise,
one must specify one or more limb-darkening coefficients for each band
of observation (Albrow et al.\ 1999a, 2000a; Afonso et al.\ 2000).
Finally, it is possible that the binary's rotation is detectable
in which case one or more parameters are required to describe that
(Dominik 1998a; Albrow et al.\ 2000a).  In addition, binary light curves often
have data from several observatories in which case one needs two parameters
(source flux and background flux) for each observatory.  

\subsection{Binary Lens Lightcurve Fitting}

	The problem of fitting binary-lens lightcurves is extremely
complicated and is still very much under active investigation.  
There are actually three inter-related difficulties.  First, as discussed
in \S\ 4.1, the parameter space is large.  Second, it turns out the
minima of $\chi^2$ over this space are not well behaved.  Third, the evaluation
of the magnification for a finite source straddling or near a caustic can
be computationally time consuming.  The combination of these three factors
means that a brute force search for solutions can well fail or, worse yet,
settle on a false minimum.

	The ideas for tackling these problems go back to the detection of
the first binary microlens OGLE-7 (Udalski et al.\ 1994).  There are two
major categories: ideas for improving efficiency in the evaluation of
the magnification, and ideas for cutting down the region of parameter space 
that must be searched.  For the first, various methods have been developed
by Kayser \& Schramm (1988), Bennett \& Rhie (1996), Gould \& Gaucherel (1997),
Wambsganss (1997) and Dominik (1998b), although in fact all are still fairly
time consuming.  For the second, Mao \& Di Stefano (1995) developed a densely 
sampled library of point-source binary microlensing events, each of which is 
characterized by cataloged ``features'' such as the number of maxima,
heights of peaks, etc.  They can then examine individual 
events, characterize their ``features,'' and search their library for
events that are consistent with these features.  Di Stefano \& Perna (1997)
suggested that binary lenses could be fitted by decomposing the observed light 
curve into a linear combination of basis functions.  The coefficients of
these functions could then be compared to those fitted to a library of events 
in order to isolate viable regions of parameter space.  This is essentially the
same method as Mao \& Di Stefano (1995), except that, rather than use gross 
features to identify similar light curves, one uses the coefficients of the 
polynomial expansion, which is more quantitative and presumably more robust.

	Albrow et al.\ (1999b) developed a hybrid approach that both simplifies
the search of parameter space and vastly reduces the computation time for
individual light curves.  It makes use of the fact that one of the very 
few things that is simple about a binary microlens
is the behavior of its magnification very near to a caustic.  A source inside 
a caustic will be imaged into five images, while outside the caustics it will
be imaged into three images.  Hence, at the caustic two images appear or
disappear.  These images are infinitely magnified.  In the immediate 
neighborhood of a caustic (assuming one is not near a cusp), the magnification
of the two new images diverges as $A_2 \propto (-\Delta u_\perp)^{-1/2}$,
where $\Delta u_\perp$ is the perpendicular separation of the source from
the caustic (in units of $\theta_{\rm E}$).  On the other hand, the
three other images are unaffected by the approach of the caustic, so
$A_3\sim$const.  Hence, the total magnification is given by 
(Schneider \& Weiss 1987)
\begin{equation}
A = A_2 + A_3 \simeq \biggl(-{\Delta u_\perp\over u_r}\biggr)^{-1/2} 
\Theta(-\Delta u_\perp) + A_{cc},
\end{equation}
where $u_r$ is a constant that characterizes the approach to the caustic,
$A_{cc}$ is the magnification just outside the caustic crossing, and
$\Theta$ is a step function.  For a source of uniform brightness, or
limb darkened in some specified way, one can therefore write a relatively
simple expression for the total magnification as a function $\Delta u_\perp$
(Albrow et al.\ 1999b; Afonso et al.\ 2000).  

\subsection{Degeneracies}

	By fitting just the caustic-crossing data to a simple form based
on equation (16), Albrow et al.\ (1999b) were able to reduce the search
space from 7 to 5 dimensions and so effect a brute force search.  This
turned up a degeneracy between a wide-binary geometry and a close-binary 
geometry that both equally well accounted for the observed light curve.
The Albrow et al.\ (1999b) data did not cover large parts of the light curve,
but even when Afonso et al.\ (2000) combined these data with very extensive
data sets from four other microlensing collaborations, the close/wide
degeneracy survived.  Simultaneously, Dominik (1999b) discovered a whole
class of close/wide binary degeneracies whose roots lie deep in the lens
equation itself.  

	This was unexpected.  It was previously known that various observed
light curves could be fit by several radically different binary geometries
(Dominik 1999a).  However, Han et al.\ (1999) showed that these geometries
produced radically different astrometric deviations, so that the photometric
degeneracy could be taken to be in some sense ``accidental''.  Moreover,
for all of the examples examined by Dominik (1999a) and Han et al.\ (1999),
the data, while reasonably good, were substantially poorer in quality
than those used by Afonso et al.\ (2000).  Hence, it was plausible to hope
that with better photometric data, the degeneracies could be resolved.
However, since the Dominik (1999b) degeneracies are rooted in the lens
equation, they may prove more intractible.  For example, Gould \& Han (2000)
showed that, in contrast to the cases examined by Han et al.\ (1999),
the wide and close models presented by Afonso et al.\ (2000) generated
similar astrometric deviations, although they could be distinguished with
sufficiently late time data.  On the other hand, Albrow et al.\ (2000b)
showed that in at least one case this degeneracy is easily broken with
photometric data alone.

\section{Femtolensing}

	Femtolensing refers to interference effects in microlensing and
derives its name from the very small angular scales that are usually 
required to produce such effects (Gould 1992a).  To date, work on 
femtolensing has been almost entirely theoretical, although there has
been at least one significant observational result.

	As I mentioned in \S\ 2.2, the magnification is given by the ratio 
of the area of the image to the area of the source, but this applies only
to single images.  Multiple images will in general interfere with one
another.  The reason that this can generally be ignored (and so one can
write $A=A_++A_-$ as I did in eq.\ 6) is that usually the source is large
enough that for some parts the two images interfere constructively, and for
other parts they do so destructively, so that 
one can simply add intensities rather than amplitudes.

	The validity of this approximation then depends on how rapidly
the relative phase (proportional to the time delay) varies across the source.
For a point lens and for $u\ll 1$, the time delay between the images is
given by $\Delta t = 8 G M/c^3 (1+z_L)u$ where $z_L$ is the redshift of
the lens.  The phase delay is therefore,
\begin{equation}
\Delta \phi = {E\Delta t\over h} = 9.5\times 10^{9} 
{M\over M_\odot}\,{E\over \rm eV}\,(1+z_L)u,
\end{equation}
where $E$ is the energy of the photon.  Hence, for ordinary Galactic
microlensing observed in optical light, the interference effects are completely
wiped out unless the source covers only a tiny part of the Einstein ring,
$\rho_*\la 10^{-10}$.  Thus, applications of femtolensing require a search
for unusual regions of parameter space.  However, if such regions of
parameter space can be identified, the effects can be dramatic:  if the
magnifications of the images are written in terms of $u_\pm$ (eq.\ 6), it
is not difficult to show that,
\begin{equation}
{\cal A}_\pm = (A_+^{1/2} \pm A_-^{1/2})^2 = (1 + 4/u^2)^{\pm 1/2},
\end{equation}
where ${\cal A}_\pm$ are the magnifications at constructive and destructive
interference.  Hence the ratio of the peaks to the troughs is
${\cal A}_+/{\cal A}_-=(1+4/u^2)$, which for typical 
$u\sim 0.5$
can be quite large.  Hence, the interference pattern should manifest itself
regardless of what intrinsic spectral features the source has.

	Femtolensing was first discussed by Mandzhos (1981) and further
work was done by Schneider \& Schmidt-Burgk (1985) and Deguchi \& Watson
(1986).  Peterson \& Falk (1991) were the first to confront the
problem of scales posed by equation (17).  They considered radio sources
(and so gained about 5 orders of magnitude relative to the optical) and
advocated high signal-to-noise ratio observations that could detect the
{\cal O}(1/N) effects if the source subtends $N$ fringes.  

	Gould (1992a) sought to overcome the huge factor in equation (17) by
going to smaller $M$.  He showed that asteroid-sized objects 
($M\sim 10^{-16}\,M_\odot$) could femtolens gamma-ray bursts 
(i.e., $E\sim\,$MeV).  Plugging these numbers into equation (17) yields
phase changes $\Delta\phi\sim 1$ over the entire Einstein ring.  This
is the only femtolensing suggestion that has ever been carried out in
practice: Marani et al.\ (1999) searched BATSE and Ulysses data for
femtolensing (as well as several other types of lensing) and used their
null results to place weak limits on cosmological lenses in this mass
regime.  Kolb \& Tkachev (1996) showed that these type of observations
can also be used to probe for axion clusters to determine if such axions
make up the dark matter.

	Ulmer \& Goodman (1995) developed a formalism capable of going
beyond the semi-classical approximation of previous investigations.
They thereby found effects that are in principle observable even at optical 
wavelegths and solar masses, despite the seemingly pessimistic implications
of equation (17). Jaroszynski, \& Paczy\'nski (1995) then showed that
these results could have implications for microlensing observations of
Huchra's Lens.

	I close this review with a description of another potential
application of femtolensing due to Gould \& Gaudi (1997).  The Einstein
ring associated with a typical M star at $\sim 20\,$pc is 
$\theta_{\rm E}\sim 10\,$mas, corresponding to $r_{\rm E}\sim 0.2\,$AU.
Hence, the binary companions of such stars are likely to be several
orders of magnitude farther away.  The binary lens can then be approximated
as a point lens perturbed by a weak shear due to the companion.  This
produces a Chang-Refsdal lens (Chang \& Refsdal 1979, 1984), 
which is more thoroughly described by Schneider, Ehlers, \& Falco (1992).
A source lying inside the caustic and near one of the cusps will produce
four images near the Einstein ring, three highly magnified images on the same
side of the Einstein ring and one moderately magnified image (which we will
henceforth ignore) on the other.  Depending on the details of the geometry,
the three images can easily be magnified $10^6$ times in one direction
but are not much affected (indeed shrunk by a factor 2) in the other.
Thus a $10^8\,M_\odot$ black hole (Schwarzschild radius 2 AU) at the center
of a quasar at 1 Gpc, could be magnified in one direction from 2 nas, to 2 
mas, and so could easily be
resolved with a space-based optical interferometer having a baseline of a few
hundred meters.  

	The only problem, then, is how to get similar resolution in the
other direction.  Femtolensing provides the answer.  Because the three images
are not magnified in the direction perpendicular to the Einstein ring, they
will each necessarily contain images not only of the black hole, but of 
considerable other junk along one-dimensional bands cutting through its 
neighborhood.  The three bands from the three images will cut accross
one another and will only intersect along a length approximately equal to
the width of the bands, i.e., $\sim 1\,$AU.  If light from two of these
images is brought together and dispersed in a spectrograph, then only
the light from the intersecting region will give rise to interference
fringes.  The amplitude of these fringes will be set by the ratio of the
length in the source plane over which the time delay between the images 
differs by 1 wavelength to the width of the bands.  For typical parameters
this could be a few percent.  Since the $V\sim 22$ quasar will be magnified
$\sim 10^6$ times to $V\sim 7$, this should not be difficult to detect.

	There are a few difficulties that must be overcome to make this
work in practice.  First, the nearest position from which a dwarf star
appears aligned with a quasar lies about 40 AU from the Sun.  So while
the huge ``primary'' of this ``femtolens telescope'' (the dwarf star) comes
for free, getting the ``secondary optics'' aligned with the primary is
a big job.  Second, unlike other space missions that deliver payloads to
40 AU (e.g., the Voyagers), this package must be stopped at 40 AU so
that it remains aligned with the dwarf star and quasar.  Third, there
are station keeping problems because the telescope will gradually fall
out of alignment due to the Sun's gravity and will have to be realigned
about every 10 hours.  However, what sort of theorist would recoil from
a few engineering challenges?

\section{Conclusion}

	When microlensing experiments began in the early 1990s, few 
participants expected that there was much room for theoretical
development at all.  The physical effect (eq.\ 1) was completely understood,
and the basic equations of microlensing had all been worked out.  A decade
later, microlensing theory has shown itself to be a very dynamic field.
The original problems turned out to be much richer than expected, while
new observations and new developments in instrumention have raised new
problems.  Thus, microlensing theory promises to remain vibrant.

\noindent{\bf Acknowledgements}: 
This work was supported by grant AST 97-27520 from the NSF.


\begin{references}
\reference Afonso, C.\ et al.\ 2000, \apj, 532, 000 (astro-ph/9907247) 
\reference Albrow, M.\ et al.\ 1999a, \apj, 522, 1011 
\reference Albrow, M.\ et al.\ 1999b, \apj, 522, 1022 
\reference Albrow, M.\ et al.\ 2000a, \apj, 534, 000 (astro-ph/9910307)
\reference Albrow, M.\ et al.\ 2000b, in preparation
\reference Alcock, C., et al.\ 1995 \apj, 454, L125 % parallax
\reference Alcock, C., et al.\ 1997 \apj, 491, 436 % 95-30
\reference Alcock, C., et al.\ 2000 \apj, submitted (astro-ph/9907369)
\reference Bennett, D.B.\ 1997, BAAS, 191, 8303
\reference Bennett, D.P.\ \& Rhie S.H.\ 1996, \apj, 472, 660
\reference Boden, A.F., Shao, M., \& Van Buren, D.\ 1998 \apj, 502, 538
\reference Boutreux, T., \& Gould, A.\ 1996, \apj, 462, 705
\reference Chang, K.\ \& Han, C.\ 1999, \apj, 525, 434 (astro-ph/00011930)
\reference Chang, K.\ \& Refsdal, S.\ 1979, Nature, 282, 561
\reference Chang, K.\ \& Refsdal, S.\ 1984, \aap, 130, 157
\reference Deguchi, S., \& Watson, W.D.\ 1986, \apj, 307, 30
\reference Di Stefano, R., \& Perna, R.\ 1997, \apj, 488, 55
\reference Dominik, M.\ 1998a, \aap, 329, 361
\reference Dominik, M.\ 1998b, \aap, 333, L79
\reference Dominik, M.\ 1999a, \aap, 341, 943
\reference Dominik, M.\ 1999b, \aap, 349, 108
\reference Einstein, A.\ 1936, Science, 84, 506
\reference Froeschle, M., Mignard, F., \& Arenou, F.\ 1997, Proceedings of 
the ESA Symposium 'Hipparcos -- Venice '97',  p.\ 49, ESA SP-402
\reference Gaudi, B.S., \& Gould, A.\ 1997, 477, 152
\reference Gaudi, B.S., Naber, R.M., \& Sackett, P.D.\ 1998, 502, L33
\reference Gould, A.\ 1992a, \apj, 386, L5
\reference Gould, A.\ 1992b, \apj, 392, 442
\reference Gould, A.\ 1994a, \apj, 421, L71
\reference Gould, A.\ 1994b, \apj, 421, L75
\reference Gould, A.\ 1995, \apj, 441, L21
\reference Gould, A.\ 1996, \pasp, 108, 465
\reference Gould, A.\ 1997, \apj, 480, 188
\reference Gould, A.\ 2000a, \apj, submitted (astro-ph/9909455)
\reference Gould, A.\ 2000b, \apj, submitted (astro-ph/0001421)
\reference Gould, A.\ \& Andronov, N.\ 1999, \apj, 516, 236
\reference Gould, A.\ \& Gaucherel, C.\ 1997, \apj, 477, 580
\reference Gould, A.\ \& Gaudi, B.S.\ 1997, \apj, 486, 687
\reference Gould, A.\ \& Han, C.\ 2000, \apj, 538, 000 (astro-ph/00011930)
\reference Gould, A., \& Salim, S.\ 1999, \apj, 524, 794
\reference Griest, K.\ 1991, \apj, 366, 412
\reference Han, C., Chun, M.S., \& Chang, K.\ 1999, 526, 405
\reference Han, C., \& Gould, A.\ 1997, \apj, 480, 196
\reference Han, C., \& Jeong, Y.\ 1999, \mnras, 309, 404
\reference Han, C., \& Kim, H.-I.\ 2000, \apj, 528, 687
\reference Han, C., \& Kim, T.-W.\ 1999, \mnras, 305, 795
\reference Hardy, S.J., \& Walker, M.A.\ 1995, \mnras, 276, L79
\reference Holz, D.E., \& Wald, R.M.\ 1996, ApJ, 471, 64 
\reference Honma, M.\ 1999, \apj, 517, L35
\reference Jaroszynski, M., \& Paczy\'nski, B.\ 1995, \apj, 455, 433
\reference Jeong, Y., Han, C., \& Park, S.H.\ 1999, \mnras, 304, 845
\reference Kayser, R., \& Schramm, T.\ 1998, \aap, 191, 39
\reference Kolb, E.W., \& Tkachev, I.I.\ 1996, \apj, 460, L25
\reference Mandzhos, A.V.\ 1981, \sovast, 7, L213
\reference Mao, S.\ 1999a, \aap, 350, L19
\reference Mao, S.\ 1999b, in Gravitational Lensing, Recent Progress and Future
Goals, Boston University, July 1999, ed.\ T.G.\ Brainerd and C.S.\ Kochanek,
in press (astro-ph/9909302)
\reference Mao, S., \& Di Stefano, R.\ 1995, \apj, 440 22
\reference Marani, G.F., Nemiroff, R.J., Norris, J.P., Hurley, K., \& 
Bonnell, J.T., 1999, \apj 512, L13
\reference Miralda-Escud\'e, J.\ 1996, \apj, 470, L113
\reference Nemiroff, R.J.\ 1991, \aap, 247, 73
\reference Nemiroff, R.J.\ \& Wickramasinghe, W.A.D.T.\ 1994, \apj, 424, L21
\reference Paczy\'nski, B.\ 1986, \apj, 304, 1
\reference Paczy\'nski, B.\ 1995, Acta Astron., 45, 345
\reference Paczy\'nski, B.\ 1996, \araa, 34, 419
\reference Paczy\'nski, B.\ 1998, \apj, 494, L23
\reference Peterson, J.B., \& Falk, T.\ 1991, \apj, 374, L5
\reference Refsdal, S.\ 1964, \mnras, 128, 295
\reference Refsdal, S.\ 1966, \mnras, 134, 315
\reference Rhie, S.H.\ 1997, \apj, 484, 63
\reference Roulet, E., \& Mollerach, S.\ 1997, Physics Reports, 279, 68
\reference Safizadeh, N., Dalal, N., \& Griest, K.\ 1999, \apj, 522, 512
\reference Salim, S., \& Gould, A., 2000 \apj, 539, 000
\reference Schneider, P., Ehlers, J., \& Falco, E.E.\ 1992, 
Gravitational Lenses (Berlin: Springer)
\reference Schneider, P., \& Schmidt-Burgk, J.\ 1985, \aap, 148, 369
\reference Schneider, P., \& Weiss, A. 1986, A\&A, 164, 237
\reference Schneider, P., \& Weiss, A. 1987, A\&A, 171, 49
\reference Udalski, A., Szyma\'nski, M., Pietrzy\'nski, G., Kubiak, M.,  
Wo\'zniak, P., \& \.Zebru\'n, K.\ 1998, Acta Astron., 48, 431
\reference Ulmer, A., \& Goodman, J.\ 1995, \apj 442, 67
\reference van Belle, G.T.\ 1999, \pasp, 111, 1515
\reference Wambsganss, J.\ 1997, \mnras, 284, 172
\reference Witt, H.J.\ 1990, \aap, 236, 311
\reference Witt, H.J.\ \& Mao, S.\ 1994, \apj, 429, 66

\end{references}
\end{document}